\documentclass[aps,prd,twocolumn,groupedaddress]{revtex4-2}
\usepackage{graphicx}
\usepackage{hyperref}
\begin{document}
\title{Implications of the detection of sub-PeV diffuse $\gamma$ rays from the Galactic disk \\ apart from discrete sources}
\author{T.A. Dzhatdoev}
\email[]{timur1606@gmail.com}
\affiliation{Federal State Budget Educational Institution of Higher Education, M.V. Lomonosov Moscow State University, Skobeltsyn Institute of Nuclear Physics (SINP MSU), 1(2), Leninskie gory, GSP-1, 119991 Moscow, Russia}
\affiliation{Institute for Nuclear Research of the Russian Academy of Sciences, 60th October Anniversary Prospect 7a, Moscow 117312, Russia}
\affiliation{Institute for Cosmic Ray Research, University of Tokyo, 5-1-5 Kashiwanoha, Kashiwa, Japan}
\date{\today}
\begin{abstract}
Very recently, the Tibet-AS$\gamma$ collaboration reported the detection of $\gamma$ rays from the galactic disk in the energy range of 100 TeV -- 1 PeV. Remarkably, many of these $\gamma$ rays  were observed apart from known very high energy ($E>100$ GeV) $\gamma$-ray sources. These results are best understood if these diffuse $\gamma$ rays: 1) were produced by a conventional rather than an exotic (i.e. dark matter decay or annihilation) process, 2) have a hadronic rather than a leptonic origin, 3) were produced in impulsive rather than stable sources or, alternatively, in optically thick sources. In addition to that, the detection of the sub-PeV diffuse $\gamma$ rays implies a limit on the flux of neutrinos from the Galactic disk and a lower limit on the rigidity of the cutoff in the Galactic cosmic ray spectrum.
\end{abstract}
\maketitle
\section{Introduction}

Galactic cosmic rays interact with gas and radiation fields inside the sources and in the Galactic volume, producing $\gamma$ rays, electrons, positrons, and neutrinos \cite{Hayakawa1952,Ginzburg1963,Pollack1963,Ginzburg1965,Stecker1970}. $\gamma$ rays and neutrinos travel in straight lines, allowing the observer to discern their source(s). Arrival directions of high energy ($E > 100$ MeV) $\gamma$ rays from Galactic sources are concentrated towards the Galactic plane \cite{Clark1968,Kraushaar1972,Fichtel1975,Mayer1982,Hunter1997,Ackermann2012,Acero2016}. A part of Galactic $\gamma$ rays is ``diffuse'', i.e. these particles are observed apart from ``discrete'' (point-like or slightly extended) sources.

$\gamma$ rays of very high ($E > 100$ GeV) and super high energy ($E > 100$ TeV) may be detected with ground-based installations such as imaging atmospheric Cherenkov telescopes (IACT) \cite{Hinton2004,Aleksic2016,Krennrich2004,Acharya2013} and air shower arrays (e.g. \cite{Apel2017,Amenomori2019,Aartsen2020}). Very recently, the Tibet-AS$\gamma$ collaboration reported the discovery of diffuse $\gamma$ rays concentrating towards the Galactic plane \cite{Amenomori2021} (hereafter A21). This observation has a number of interesting and important theoretical implications, some of which are considered below. In particular:\\
1. a conventional (astrophysical) production mechanism of these $\gamma$ rays is favoured over an exotic mechanism (i.e. from dark matter decay or annihilation) (Sect.~\ref{sect:DarkMatter})\\
2. the hadronic production mechanism is more likely than the leptonic one (Sect.~\ref{sect:Mechanism})\\
3. the high fraction of $\gamma$ rays detected apart from discrete sources implies that the cosmic ray acceleration sites are either optically thick to these $\gamma$ rays or that these accelerators were more active in the past than now (Sect.~\ref{sect:Sources})\\
4. galactic cosmic ray models with a very low energy of the proton ``knee'' are excluded if the change in the spectral index of elemental spectra is large enough (Sect.~\ref{sect:Knee}).

In addition, we note that diffuse Galactic $\gamma$-rays may help constraining the Galactic component of IceCube neutrinos (e.g. \cite{Ahlers2014}).

\section{Conventional or exotic production mechanism? \label{sect:DarkMatter}}

Using the model of \cite{Lipari2018} (hereafter LV18) assuming the production of diffuse $\gamma$ rays by cosmic rays in hadronuclear interactions, A21 show that their data are reasonably well approximated with the LV18 model. However, one could speculate that the flux of $\gamma$ rays reported in A21 could be produced by decay or annihilation of dark matter particles. In this section we assume that the large-scale distribution of Galactic dark matter follows the Navarro-Frenk-White (NFW) density distribution \cite{Navarro1997}.

\begin{figure}
\vspace{0.2cm}
\includegraphics[width=9cm]{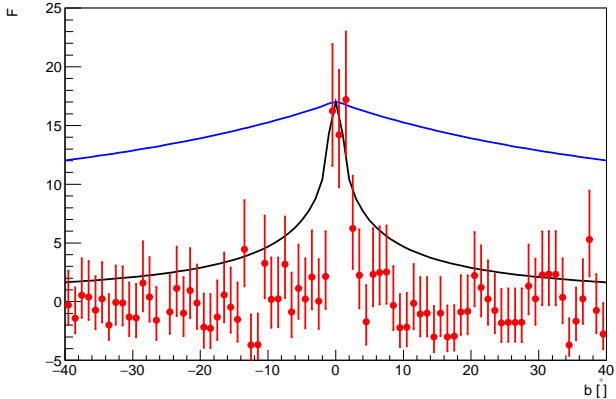}
\caption{The Galactic latitude distribution of excess $\gamma$-ray-like events according to A21 for the 158--398~TeV energy bin (red histogram) in comparison with the model angular profiles for the case of dark matter decay (blue curve) and annihilation (black curve).
\label{fig:DarkMatter}}
\end{figure}

LV18 proposed a test of dark matter origin for Galactic diffuse $\gamma$ rays using their distribution on the Galactic latitude (see Fig. 17 of LV18 and associated text). Following the approach of LV18, we calculated the angular distributions for the case of dark matter annihilation and decay and compare these with data presented in A21 for the 158--398~TeV energy bin (Fig.~\ref{fig:DarkMatter}). Here, for simplicity, the effects of non-uniform sky exposure of the Tibet-AS$\gamma$ array and $\gamma$-ray absorption in the Galaxy \cite{Moskalenko2006,Vernetto2016,Porter2018} were neglected. Estimates show that the proper account of the exposure non-uniformity and the $\gamma$-ray absorption result in a broadening of the latitude distribution. 

The decay model poorly fits the data: the resulting latitude distribution is far too broad. Even for annihilating dark matter, this distribution does not provide a good fit to the data. Moreover, the annihilation model is less attractive in view of the unitarity limit on the mass of dark matter particle \cite{Griest1990}. Detailed constraints on dark matter decay time / annihilation cross section are in preparation and will be published elsewhere.

\section{Hadronic or leptonic $\gamma$ rays? \label{sect:Mechanism}}

Cosmic rays excite turbulence in the interstellar medium, inhibiting the cosmic ray transport outside of their sources \cite{Abeysekara2017}. Assuming the diffusion coefficient according to eq. (3) of \cite{Johannesson2019} with $r_{z} = 10$ pc, $r_{t} = 100$ pc, $\beta = 1$, $\delta = 0.35$, $R_{0} = 4$ GV, $D_{0} = 4.0\times10^{28}$ cm$^{2}$/s, $D_{z} = D_{0}/100$, we estimate the typical time needed to travel the central 20 pc as $\sim 100$ years (this time is somewhat greater for the greater radius of 100 pc, $\sim 200$ years). The typical synchrotron cooling time for electrons is $\approx 2(B/ 100 \mu G)^{-2}(E_{e}/ 500 TeV)^{-1}$ years (e.g. \cite{Albert2021}), i.e. about 100 years for $E_{e} = 500$ TeV and $B = 15 \mu$G. We conclude that for the typical distance to the source in excess of 1 kpc these electrons would be confined inside a 1$^{\circ}$ circle as seen by a distant observer, resulting in a very sharp concentration of $\gamma$-rays near discrete sources, in stark contradiction to the results of A21 \footnote{the only evident exception is the Cygnus Cocoon region}. We note that a similar qualitative argument was put forward in A21, without, however, quantitative estimates. Additional constraints could be obtained from the balance of energy gain and losses during the acceleration process.

\section{The nature of cosmic ray sources \label{sect:Sources}}

Now consider the escape of protons and nuclei from the sources. The typical escape time is $\sim100$ years (see the previous section). The typical acceleration time up to the knee \cite{Kulikov1959,Fowler2001,Aglietta2004,Antoni2005,Aartsen2013} $t_{acc}\sim D/v_{s}^{2}= (cE)/(3eBv_{s}^{2})$ ($v_{s}$ is the shock front velocity). For stable Galactic hadronic PeVatrons such as star forming regions \cite{Montmerle1979,Casse1980,Cesarsky1983,Bykov2014,Aharonian2019} is $t_{acc}\sim10^{3}$ years or even more. 

The typical lifetime of 3 PeV cosmic rays in the Galactic volume is $\sim 5\times10^{4}$ years (e.g. \cite{Lipari2019}). The typical contrast of gas densities between the sources and the Galactic volume is about $10^{2}-10^{3}$. The number of produced $\gamma$ rays is proportional to the concentration of the gas and the time spent inside particular regions (i.e. inside the discrete sources and inside the Galactic disc, but outside the discrete sources). We conclude that the time spent in sources should be less than several hundred years in order to not overproduce $\gamma$-rays near the discrete sources, in stark contrast to the above estimates. We conclude that the sources are likely to be impulsive or optically thick for $>100$~TeV $\gamma$ rays.




\section{Cosmic-ray knee \\ constrained with $\gamma$ rays \label{sect:Knee}}

\begin{figure}
\vspace{0.2cm}
\includegraphics[width=9cm]{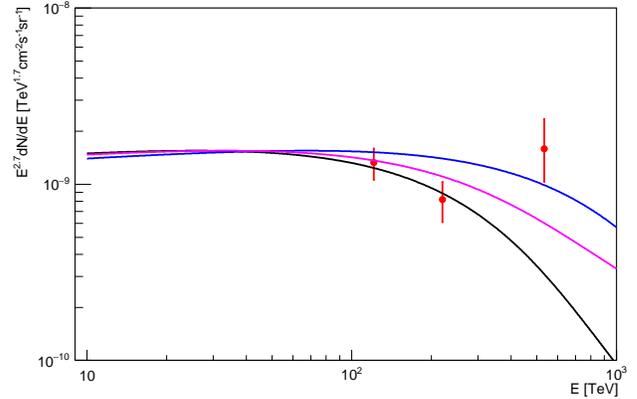}
\caption{The knee in $\gamma$-rays for several sets of model parameters compared to the spectrum of diffuse $\gamma$ rays measured with Tibet-AS$\gamma$ (more details in the text). \label{fig:Knee}}
\end{figure}

The spectrum of $\gamma$-rays measured with the Tibet-AS$\gamma$ array together with several model curves is shown in Fig.~\ref{fig:Knee}. For model curves, the primary proton spectrum was assumed to follow eq. (2) of \cite{Hoerandel2003}. Only primary protons were considered. Black curve corresponds to the proton spectral index below the knee $\gamma_{1} = 2.7$, $\Delta\gamma = 2$, the energy of the knee $E_{br} = 1$ PeV and $\epsilon_{c} = 10$. Blue curve is for the same parameters, except $E_{br} = 3$ PeV, magnenta curve is for the same parameters as black curve, except $\Delta\gamma = 1$. Remarkably, results for smaller $\epsilon_{c}$ down to 1 are similar to those presented in the graph. We conclude that relatively small values of $E_{br} < 1$ PeV are excluded for sufficiently large values of $\Delta\gamma$. We note that much better constraints could likely be achieved using the data of the LHAASO experiment \cite{Guo2014}.


\section{Conclusions \label{sect:Conclusions}}

The discovery of diffuse superhigh energy $\gamma$-rays with Tibet-AS$\gamma$ opened a new area of study in $\gamma$-ray astronomy, capable of constraining dark matter properties, probing the Galactic neutrino component, and unveiling the nature of cosmic ray sources. New data are expected from the LHAASO experiment shortly \cite{LHAASO-arxiv-2019}. Directions around $\gamma$ rays registered with Tibet-AS$\gamma$ (and, hopefully, LHAASO) could be studied with existing IACT arrays H.E.S.S., MAGIC, VERITAS, as well as with the forthcoming CTA array \cite{Actis2011,Acharya2013} in order to put further constrain on the possible contribution from discrete sources to the diffuse $\gamma$-ray flux.

\begin{acknowledgments}
The author is grateful to Prof.~P.~Lipari and Dr.~S.~Vernetto for sharing their model of $\gamma$-ray absorption in the Galaxy (Ref. [24]). Helpful discussions with Prof.~I.V.~Moskalenko and Prof.~ S.V.~Troitsky are gratefully acknowledged. This work is supported in the framework of the State project ``Science'' by the Ministry of Science and Higher Education of the Russian Federation under the contract 075-15-2020-778. All graphs in the present paper were produced with the ROOT software toolkit \cite{Brun1997}. This research has made use of the NASA ADS bibliographical system.
\end{acknowledgments}
\bibliography{Gamma-Diffuse}

\begin{thebibliography}{47}%
\makeatletter
\providecommand \@ifxundefined [1]{%
 \@ifx{#1\undefined}
}%
\providecommand \@ifnum [1]{%
 \ifnum #1\expandafter \@firstoftwo
 \else \expandafter \@secondoftwo
 \fi
}%
\providecommand \@ifx [1]{%
 \ifx #1\expandafter \@firstoftwo
 \else \expandafter \@secondoftwo
 \fi
}%
\providecommand \natexlab [1]{#1}%
\providecommand \enquote  [1]{``#1''}%
\providecommand \bibnamefont  [1]{#1}%
\providecommand \bibfnamefont [1]{#1}%
\providecommand \citenamefont [1]{#1}%
\providecommand \href@noop [0]{\@secondoftwo}%
\providecommand \href [0]{\begingroup \@sanitize@url \@href}%
\providecommand \@href[1]{\@@startlink{#1}\@@href}%
\providecommand \@@href[1]{\endgroup#1\@@endlink}%
\providecommand \@sanitize@url [0]{\catcode `\\12\catcode `\$12\catcode
  `\&12\catcode `\#12\catcode `\^12\catcode `\_12\catcode `\%12\relax}%
\providecommand \@@startlink[1]{}%
\providecommand \@@endlink[0]{}%
\providecommand \url  [0]{\begingroup\@sanitize@url \@url }%
\providecommand \@url [1]{\endgroup\@href {#1}{\urlprefix }}%
\providecommand \urlprefix  [0]{URL }%
\providecommand \Eprint [0]{\href }%
\providecommand \doibase [0]{https://doi.org/}%
\providecommand \selectlanguage [0]{\@gobble}%
\providecommand \bibinfo  [0]{\@secondoftwo}%
\providecommand \bibfield  [0]{\@secondoftwo}%
\providecommand \translation [1]{[#1]}%
\providecommand \BibitemOpen [0]{}%
\providecommand \bibitemStop [0]{}%
\providecommand \bibitemNoStop [0]{.\EOS\space}%
\providecommand \EOS [0]{\spacefactor3000\relax}%
\providecommand \BibitemShut  [1]{\csname bibitem#1\endcsname}%
\let\auto@bib@innerbib\@empty
\bibitem [{\citenamefont {Hayakawa}(1952)}]{Hayakawa1952}%
  \BibitemOpen
  \bibfield  {author} {\bibinfo {author} {\bibfnamefont {S.}~\bibnamefont
  {Hayakawa}},\ }\href {https://doi.org/10.1143/ptp/8.5.571} {\bibfield
  {journal} {\bibinfo  {journal} {Progress of Theoretical Physics}\ }\textbf
  {\bibinfo {volume} {8}},\ \bibinfo {pages} {571} (\bibinfo {year}
  {1952})}\BibitemShut {NoStop}%
\bibitem [{\citenamefont {{Ginzburg}}\ and\ \citenamefont
  {{Syrovatskii}}(1963)}]{Ginzburg1963}%
  \BibitemOpen
  \bibfield  {author} {\bibinfo {author} {\bibfnamefont {V.~L.}\ \bibnamefont
  {{Ginzburg}}}\ and\ \bibinfo {author} {\bibfnamefont {S.~I.}\ \bibnamefont
  {{Syrovatskii}}},\ }\href@noop {} {\emph {\bibinfo {title} {{The Origin of
  Cosmic Rays}}}}\ (\bibinfo {year} {1963})\BibitemShut {NoStop}%
\bibitem [{\citenamefont {Pollack}\ and\ \citenamefont
  {Fazio}(1963)}]{Pollack1963}%
  \BibitemOpen
  \bibfield  {author} {\bibinfo {author} {\bibfnamefont {J.~B.}\ \bibnamefont
  {Pollack}}\ and\ \bibinfo {author} {\bibfnamefont {G.~G.}\ \bibnamefont
  {Fazio}},\ }\href {https://doi.org/10.1103/physrev.131.2684} {\bibfield
  {journal} {\bibinfo  {journal} {Physical Review}\ }\textbf {\bibinfo {volume}
  {131}},\ \bibinfo {pages} {2684} (\bibinfo {year} {1963})}\BibitemShut
  {NoStop}%
\bibitem [{\citenamefont {Ginzburg}\ and\ \citenamefont
  {Syrovatski{\u{\i}}}(1965)}]{Ginzburg1965}%
  \BibitemOpen
  \bibfield  {author} {\bibinfo {author} {\bibfnamefont {V.~L.}\ \bibnamefont
  {Ginzburg}}\ and\ \bibinfo {author} {\bibfnamefont {S.~I.}\ \bibnamefont
  {Syrovatski{\u{\i}}}},\ }\href
  {https://doi.org/10.1070/pu1965v007n05abeh003658} {\bibfield  {journal}
  {\bibinfo  {journal} {Soviet Physics Uspekhi}\ }\textbf {\bibinfo {volume}
  {7}},\ \bibinfo {pages} {696} (\bibinfo {year} {1965})}\BibitemShut {NoStop}%
\bibitem [{\citenamefont {Stecker}(1970)}]{Stecker1970}%
  \BibitemOpen
  \bibfield  {author} {\bibinfo {author} {\bibfnamefont {F.~W.}\ \bibnamefont
  {Stecker}},\ }\href {https://doi.org/10.1007/bf00653856} {\bibfield
  {journal} {\bibinfo  {journal} {Astrophysics and Space Science}\ }\textbf
  {\bibinfo {volume} {6}},\ \bibinfo {pages} {377} (\bibinfo {year}
  {1970})}\BibitemShut {NoStop}%
\bibitem [{\citenamefont {Clark}\ \emph {et~al.}(1968)\citenamefont {Clark},
  \citenamefont {Garmire},\ and\ \citenamefont {Kraushaar}}]{Clark1968}%
  \BibitemOpen
  \bibfield  {author} {\bibinfo {author} {\bibfnamefont {G.~W.}\ \bibnamefont
  {Clark}}, \bibinfo {author} {\bibfnamefont {G.~P.}\ \bibnamefont {Garmire}},\
  and\ \bibinfo {author} {\bibfnamefont {W.~L.}\ \bibnamefont {Kraushaar}},\
  }\href {https://doi.org/10.1086/180252} {\bibfield  {journal} {\bibinfo
  {journal} {The Astrophysical Journal}\ }\textbf {\bibinfo {volume} {153}},\
  \bibinfo {pages} {L203} (\bibinfo {year} {1968})}\BibitemShut {NoStop}%
\bibitem [{\citenamefont {Kraushaar}\ \emph {et~al.}(1972)\citenamefont
  {Kraushaar}, \citenamefont {Clark}, \citenamefont {Garmire}, \citenamefont
  {Borken}, \citenamefont {Higbie}, \citenamefont {Leong},\ and\ \citenamefont
  {Thorsos}}]{Kraushaar1972}%
  \BibitemOpen
  \bibfield  {author} {\bibinfo {author} {\bibfnamefont {W.~L.}\ \bibnamefont
  {Kraushaar}}, \bibinfo {author} {\bibfnamefont {G.~W.}\ \bibnamefont
  {Clark}}, \bibinfo {author} {\bibfnamefont {G.~P.}\ \bibnamefont {Garmire}},
  \bibinfo {author} {\bibfnamefont {R.}~\bibnamefont {Borken}}, \bibinfo
  {author} {\bibfnamefont {P.}~\bibnamefont {Higbie}}, \bibinfo {author}
  {\bibfnamefont {V.}~\bibnamefont {Leong}},\ and\ \bibinfo {author}
  {\bibfnamefont {T.}~\bibnamefont {Thorsos}},\ }\href
  {https://doi.org/10.1086/151713} {\bibfield  {journal} {\bibinfo  {journal}
  {The Astrophysical Journal}\ }\textbf {\bibinfo {volume} {177}},\ \bibinfo
  {pages} {341} (\bibinfo {year} {1972})}\BibitemShut {NoStop}%
\bibitem [{\citenamefont {Fichtel}\ \emph {et~al.}(1975)\citenamefont
  {Fichtel}, \citenamefont {Hartman}, \citenamefont {Kniffen}, \citenamefont
  {Thompson}, \citenamefont {Ogelman},\ and\ \citenamefont
  {et~al.}}]{Fichtel1975}%
  \BibitemOpen
  \bibfield  {author} {\bibinfo {author} {\bibfnamefont {C.~E.}\ \bibnamefont
  {Fichtel}}, \bibinfo {author} {\bibfnamefont {R.~C.}\ \bibnamefont
  {Hartman}}, \bibinfo {author} {\bibfnamefont {D.~A.}\ \bibnamefont
  {Kniffen}}, \bibinfo {author} {\bibfnamefont {D.~J.}\ \bibnamefont
  {Thompson}}, \bibinfo {author} {\bibfnamefont {H.}~\bibnamefont {Ogelman}},\
  and\ \bibinfo {author} {\bibnamefont {et~al.}},\ }\href
  {https://doi.org/10.1086/153590} {\bibfield  {journal} {\bibinfo  {journal}
  {The Astrophysical Journal}\ }\textbf {\bibinfo {volume} {198}},\ \bibinfo
  {pages} {163} (\bibinfo {year} {1975})}\BibitemShut {NoStop}%
\bibitem [{\citenamefont {{Mayer-Hasselwander}}\ \emph
  {et~al.}(1982)\citenamefont {{Mayer-Hasselwander}}, \citenamefont
  {{Bennett}}, \citenamefont {{Bignami}}, \citenamefont {{Buccheri}},
  \citenamefont {{Caraveo}},\ and\ \citenamefont {et~al.}}]{Mayer1982}%
  \BibitemOpen
  \bibfield  {author} {\bibinfo {author} {\bibfnamefont {H.~A.}\ \bibnamefont
  {{Mayer-Hasselwander}}}, \bibinfo {author} {\bibfnamefont {K.}~\bibnamefont
  {{Bennett}}}, \bibinfo {author} {\bibfnamefont {G.~F.}\ \bibnamefont
  {{Bignami}}}, \bibinfo {author} {\bibfnamefont {R.}~\bibnamefont
  {{Buccheri}}}, \bibinfo {author} {\bibfnamefont {P.~A.}\ \bibnamefont
  {{Caraveo}}},\ and\ \bibinfo {author} {\bibnamefont {et~al.}},\ }\href@noop
  {} {\bibfield  {journal} {\bibinfo  {journal} {Astronomy {\&} Astrophysics}\
  }\textbf {\bibinfo {volume} {105}},\ \bibinfo {pages} {164} (\bibinfo {year}
  {1982})}\BibitemShut {NoStop}%
\bibitem [{\citenamefont {Hunter}\ \emph {et~al.}(1997)\citenamefont {Hunter},
  \citenamefont {Bertsch}, \citenamefont {Catelli}, \citenamefont {Dame},
  \citenamefont {Digel},\ and\ \citenamefont {et~al.}}]{Hunter1997}%
  \BibitemOpen
  \bibfield  {author} {\bibinfo {author} {\bibfnamefont {S.~D.}\ \bibnamefont
  {Hunter}}, \bibinfo {author} {\bibfnamefont {D.~L.}\ \bibnamefont {Bertsch}},
  \bibinfo {author} {\bibfnamefont {J.~R.}\ \bibnamefont {Catelli}}, \bibinfo
  {author} {\bibfnamefont {T.~M.}\ \bibnamefont {Dame}}, \bibinfo {author}
  {\bibfnamefont {S.~W.}\ \bibnamefont {Digel}},\ and\ \bibinfo {author}
  {\bibnamefont {et~al.}},\ }\href {https://doi.org/10.1086/304012} {\bibfield
  {journal} {\bibinfo  {journal} {The Astrophysical Journal}\ }\textbf
  {\bibinfo {volume} {481}},\ \bibinfo {pages} {205} (\bibinfo {year}
  {1997})}\BibitemShut {NoStop}%
\bibitem [{\citenamefont {Ackermann}\ \emph {et~al.}(2012)\citenamefont
  {Ackermann}, \citenamefont {Ajello}, \citenamefont {Atwood}, \citenamefont
  {Baldini}, \citenamefont {Ballet},\ and\ \citenamefont
  {et~al.}}]{Ackermann2012}%
  \BibitemOpen
  \bibfield  {author} {\bibinfo {author} {\bibfnamefont {M.}~\bibnamefont
  {Ackermann}}, \bibinfo {author} {\bibfnamefont {M.}~\bibnamefont {Ajello}},
  \bibinfo {author} {\bibfnamefont {W.~B.}\ \bibnamefont {Atwood}}, \bibinfo
  {author} {\bibfnamefont {L.}~\bibnamefont {Baldini}}, \bibinfo {author}
  {\bibfnamefont {J.}~\bibnamefont {Ballet}},\ and\ \bibinfo {author}
  {\bibnamefont {et~al.}},\ }\href {https://doi.org/10.1088/0004-637x/750/1/3}
  {\bibfield  {journal} {\bibinfo  {journal} {The Astrophysical Journal}\
  }\textbf {\bibinfo {volume} {750}},\ \bibinfo {pages} {3} (\bibinfo {year}
  {2012})}\BibitemShut {NoStop}%
\bibitem [{\citenamefont {Acero}\ \emph {et~al.}(2016)\citenamefont {Acero},
  \citenamefont {Ackermann}, \citenamefont {Ajello}, \citenamefont {Albert},
  \citenamefont {Baldini},\ and\ \citenamefont {et~al.}}]{Acero2016}%
  \BibitemOpen
  \bibfield  {author} {\bibinfo {author} {\bibfnamefont {F.}~\bibnamefont
  {Acero}}, \bibinfo {author} {\bibfnamefont {M.}~\bibnamefont {Ackermann}},
  \bibinfo {author} {\bibfnamefont {M.}~\bibnamefont {Ajello}}, \bibinfo
  {author} {\bibfnamefont {A.}~\bibnamefont {Albert}}, \bibinfo {author}
  {\bibfnamefont {L.}~\bibnamefont {Baldini}},\ and\ \bibinfo {author}
  {\bibnamefont {et~al.}},\ }\href {https://doi.org/10.3847/0067-0049/223/2/26}
  {\bibfield  {journal} {\bibinfo  {journal} {The Astrophysical Journal
  Supplement Series}\ }\textbf {\bibinfo {volume} {223}},\ \bibinfo {pages}
  {26} (\bibinfo {year} {2016})}\BibitemShut {NoStop}%
\bibitem [{\citenamefont {Hinton}(2004)}]{Hinton2004}%
  \BibitemOpen
  \bibfield  {author} {\bibinfo {author} {\bibfnamefont {J.}~\bibnamefont
  {Hinton}},\ }\href {https://doi.org/10.1016/j.newar.2003.12.004} {\bibfield
  {journal} {\bibinfo  {journal} {New Astronomy Reviews}\ }\textbf {\bibinfo
  {volume} {48}},\ \bibinfo {pages} {331} (\bibinfo {year} {2004})}\BibitemShut
  {NoStop}%
\bibitem [{\citenamefont {Aleksi{\'{c}}}\ \emph {et~al.}(2016)\citenamefont
  {Aleksi{\'{c}}}, \citenamefont {Ansoldi}, \citenamefont {Antonelli},
  \citenamefont {Antoranz}, \citenamefont {Babic},\ and\ \citenamefont
  {et~al.}}]{Aleksic2016}%
  \BibitemOpen
  \bibfield  {author} {\bibinfo {author} {\bibfnamefont {J.}~\bibnamefont
  {Aleksi{\'{c}}}}, \bibinfo {author} {\bibfnamefont {S.}~\bibnamefont
  {Ansoldi}}, \bibinfo {author} {\bibfnamefont {L.}~\bibnamefont {Antonelli}},
  \bibinfo {author} {\bibfnamefont {P.}~\bibnamefont {Antoranz}}, \bibinfo
  {author} {\bibfnamefont {A.}~\bibnamefont {Babic}},\ and\ \bibinfo {author}
  {\bibnamefont {et~al.}},\ }\href
  {https://doi.org/10.1016/j.astropartphys.2015.04.004} {\bibfield  {journal}
  {\bibinfo  {journal} {Astroparticle Physics}\ }\textbf {\bibinfo {volume}
  {72}},\ \bibinfo {pages} {61} (\bibinfo {year} {2016})}\BibitemShut {NoStop}%
\bibitem [{\citenamefont {Krennrich}\ \emph {et~al.}(2004)\citenamefont
  {Krennrich}, \citenamefont {Bond}, \citenamefont {Boyle}, \citenamefont
  {Bradbury}, \citenamefont {Buckley},\ and\ \citenamefont
  {et~al.}}]{Krennrich2004}%
  \BibitemOpen
  \bibfield  {author} {\bibinfo {author} {\bibfnamefont {F.}~\bibnamefont
  {Krennrich}}, \bibinfo {author} {\bibfnamefont {I.}~\bibnamefont {Bond}},
  \bibinfo {author} {\bibfnamefont {P.}~\bibnamefont {Boyle}}, \bibinfo
  {author} {\bibfnamefont {S.}~\bibnamefont {Bradbury}}, \bibinfo {author}
  {\bibfnamefont {J.}~\bibnamefont {Buckley}},\ and\ \bibinfo {author}
  {\bibnamefont {et~al.}},\ }\href
  {https://doi.org/10.1016/j.newar.2003.12.050} {\bibfield  {journal} {\bibinfo
   {journal} {New Astronomy Reviews}\ }\textbf {\bibinfo {volume} {48}},\
  \bibinfo {pages} {345} (\bibinfo {year} {2004})}\BibitemShut {NoStop}%
\bibitem [{\citenamefont {Acharya}\ \emph {et~al.}(2013)\citenamefont
  {Acharya}, \citenamefont {Actis}, \citenamefont {Aghajani}, \citenamefont
  {Agnetta},\ and\ \citenamefont {{J. Aguilar et al.}}}]{Acharya2013}%
  \BibitemOpen
  \bibfield  {author} {\bibinfo {author} {\bibfnamefont {B.}~\bibnamefont
  {Acharya}}, \bibinfo {author} {\bibfnamefont {M.}~\bibnamefont {Actis}},
  \bibinfo {author} {\bibfnamefont {T.}~\bibnamefont {Aghajani}}, \bibinfo
  {author} {\bibfnamefont {G.}~\bibnamefont {Agnetta}},\ and\ \bibinfo {author}
  {\bibnamefont {{J. Aguilar et al.}}},\ }\href
  {https://doi.org/10.1016/j.astropartphys.2013.01.007} {\bibfield  {journal}
  {\bibinfo  {journal} {APh}\ }\textbf {\bibinfo {volume} {43}},\ \bibinfo
  {pages} {3} (\bibinfo {year} {2013})}\BibitemShut {NoStop}%
\bibitem [{\citenamefont {Apel}\ \emph {et~al.}(2017)\citenamefont {Apel},
  \citenamefont {Arteaga-Vel{\'{a}}zquez}, \citenamefont {Bekk}, \citenamefont
  {Bertaina}, \citenamefont {Bl\"{u}mer},\ and\ \citenamefont
  {et~al.}}]{Apel2017}%
  \BibitemOpen
  \bibfield  {author} {\bibinfo {author} {\bibfnamefont {W.~D.}\ \bibnamefont
  {Apel}}, \bibinfo {author} {\bibfnamefont {J.~C.}\ \bibnamefont
  {Arteaga-Vel{\'{a}}zquez}}, \bibinfo {author} {\bibfnamefont
  {K.}~\bibnamefont {Bekk}}, \bibinfo {author} {\bibfnamefont {M.}~\bibnamefont
  {Bertaina}}, \bibinfo {author} {\bibfnamefont {J.}~\bibnamefont
  {Bl\"{u}mer}},\ and\ \bibinfo {author} {\bibnamefont {et~al.}},\ }\href
  {https://doi.org/10.3847/1538-4357/aa8bb7} {\bibfield  {journal} {\bibinfo
  {journal} {The Astrophysical Journal}\ }\textbf {\bibinfo {volume} {848}},\
  \bibinfo {pages} {1} (\bibinfo {year} {2017})}\BibitemShut {NoStop}%
\bibitem [{\citenamefont {Amenomori}\ \emph {et~al.}(2019)\citenamefont
  {Amenomori}, \citenamefont {Bao}, \citenamefont {Bi}, \citenamefont {Chen},
  \citenamefont {Chen},\ and\ \citenamefont {et~al.}}]{Amenomori2019}%
  \BibitemOpen
  \bibfield  {author} {\bibinfo {author} {\bibfnamefont {M.}~\bibnamefont
  {Amenomori}}, \bibinfo {author} {\bibfnamefont {Y.}~\bibnamefont {Bao}},
  \bibinfo {author} {\bibfnamefont {X.}~\bibnamefont {Bi}}, \bibinfo {author}
  {\bibfnamefont {D.}~\bibnamefont {Chen}}, \bibinfo {author} {\bibfnamefont
  {T.}~\bibnamefont {Chen}},\ and\ \bibinfo {author} {\bibnamefont {et~al.}},\
  }\bibfield  {journal} {\bibinfo  {journal} {Physical Review Letters}\
  }\textbf {\bibinfo {volume} {123}},\ \href
  {https://doi.org/10.1103/physrevlett.123.051101}
  {10.1103/physrevlett.123.051101} (\bibinfo {year} {2019})\BibitemShut
  {NoStop}%
\bibitem [{\citenamefont {Aartsen}\ \emph {et~al.}(2020)\citenamefont
  {Aartsen}, \citenamefont {Ackermann}, \citenamefont {Adams}, \citenamefont
  {Aguilar}, \citenamefont {Ahlers},\ and\ \citenamefont
  {et~al.}}]{Aartsen2020}%
  \BibitemOpen
  \bibfield  {author} {\bibinfo {author} {\bibfnamefont {M.~G.}\ \bibnamefont
  {Aartsen}}, \bibinfo {author} {\bibfnamefont {M.}~\bibnamefont {Ackermann}},
  \bibinfo {author} {\bibfnamefont {J.}~\bibnamefont {Adams}}, \bibinfo
  {author} {\bibfnamefont {J.~A.}\ \bibnamefont {Aguilar}}, \bibinfo {author}
  {\bibfnamefont {M.}~\bibnamefont {Ahlers}},\ and\ \bibinfo {author}
  {\bibnamefont {et~al.}},\ }\href {https://doi.org/10.3847/1538-4357/ab6d67}
  {\bibfield  {journal} {\bibinfo  {journal} {The Astrophysical Journal}\
  }\textbf {\bibinfo {volume} {891}},\ \bibinfo {pages} {9} (\bibinfo {year}
  {2020})}\BibitemShut {NoStop}%
\bibitem [{\citenamefont {Amenomori}\ \emph {et~al.}(2021)\citenamefont
  {Amenomori}, \citenamefont {Bao}, \citenamefont {Bi}, \citenamefont {Chen},
  \citenamefont {Chen},\ and\ \citenamefont {et~al.}}]{Amenomori2021}%
  \BibitemOpen
  \bibfield  {author} {\bibinfo {author} {\bibfnamefont {M.}~\bibnamefont
  {Amenomori}}, \bibinfo {author} {\bibfnamefont {Y.}~\bibnamefont {Bao}},
  \bibinfo {author} {\bibfnamefont {X.}~\bibnamefont {Bi}}, \bibinfo {author}
  {\bibfnamefont {D.}~\bibnamefont {Chen}}, \bibinfo {author} {\bibfnamefont
  {T.}~\bibnamefont {Chen}},\ and\ \bibinfo {author} {\bibnamefont {et~al.}},\
  }\bibfield  {journal} {\bibinfo  {journal} {Physical Review Letters}\
  }\textbf {\bibinfo {volume} {126}},\ \href
  {https://doi.org/10.1103/physrevlett.126.141101}
  {10.1103/physrevlett.126.141101} (\bibinfo {year} {2021})\BibitemShut
  {NoStop}%
\bibitem [{\citenamefont {Ahlers}\ and\ \citenamefont
  {Murase}(2014)}]{Ahlers2014}%
  \BibitemOpen
  \bibfield  {author} {\bibinfo {author} {\bibfnamefont {M.}~\bibnamefont
  {Ahlers}}\ and\ \bibinfo {author} {\bibfnamefont {K.}~\bibnamefont
  {Murase}},\ }\bibfield  {journal} {\bibinfo  {journal} {Physical Review D}\
  }\textbf {\bibinfo {volume} {90}},\ \href
  {https://doi.org/10.1103/physrevd.90.023010} {10.1103/physrevd.90.023010}
  (\bibinfo {year} {2014})\BibitemShut {NoStop}%
\bibitem [{\citenamefont {Lipari}\ and\ \citenamefont
  {Vernetto}(2018)}]{Lipari2018}%
  \BibitemOpen
  \bibfield  {author} {\bibinfo {author} {\bibfnamefont {P.}~\bibnamefont
  {Lipari}}\ and\ \bibinfo {author} {\bibfnamefont {S.}~\bibnamefont
  {Vernetto}},\ }\bibfield  {journal} {\bibinfo  {journal} {Physical Review D}\
  }\textbf {\bibinfo {volume} {98}},\ \href
  {https://doi.org/10.1103/physrevd.98.043003} {10.1103/physrevd.98.043003}
  (\bibinfo {year} {2018})\BibitemShut {NoStop}%
\bibitem [{\citenamefont {Navarro}\ \emph {et~al.}(1997)\citenamefont
  {Navarro}, \citenamefont {Frenk},\ and\ \citenamefont {White}}]{Navarro1997}%
  \BibitemOpen
  \bibfield  {author} {\bibinfo {author} {\bibfnamefont {J.~F.}\ \bibnamefont
  {Navarro}}, \bibinfo {author} {\bibfnamefont {C.~S.}\ \bibnamefont {Frenk}},\
  and\ \bibinfo {author} {\bibfnamefont {S.~D.~M.}\ \bibnamefont {White}},\
  }\href {https://doi.org/10.1086/304888} {\bibfield  {journal} {\bibinfo
  {journal} {The Astrophysical Journal}\ }\textbf {\bibinfo {volume} {490}},\
  \bibinfo {pages} {493} (\bibinfo {year} {1997})}\BibitemShut {NoStop}%
\bibitem [{\citenamefont {Moskalenko}\ \emph {et~al.}(2006)\citenamefont
  {Moskalenko}, \citenamefont {Porter},\ and\ \citenamefont
  {Strong}}]{Moskalenko2006}%
  \BibitemOpen
  \bibfield  {author} {\bibinfo {author} {\bibfnamefont {I.~V.}\ \bibnamefont
  {Moskalenko}}, \bibinfo {author} {\bibfnamefont {T.~A.}\ \bibnamefont
  {Porter}},\ and\ \bibinfo {author} {\bibfnamefont {A.~W.}\ \bibnamefont
  {Strong}},\ }\href {https://doi.org/10.1086/503524} {\bibfield  {journal}
  {\bibinfo  {journal} {The Astrophysical Journal}\ }\textbf {\bibinfo {volume}
  {640}},\ \bibinfo {pages} {L155} (\bibinfo {year} {2006})}\BibitemShut
  {NoStop}%
\bibitem [{\citenamefont {Vernetto}\ and\ \citenamefont
  {Lipari}(2016)}]{Vernetto2016}%
  \BibitemOpen
  \bibfield  {author} {\bibinfo {author} {\bibfnamefont {S.}~\bibnamefont
  {Vernetto}}\ and\ \bibinfo {author} {\bibfnamefont {P.}~\bibnamefont
  {Lipari}},\ }\bibfield  {journal} {\bibinfo  {journal} {Physical Review D}\
  }\textbf {\bibinfo {volume} {94}},\ \href
  {https://doi.org/10.1103/physrevd.94.063009} {10.1103/physrevd.94.063009}
  (\bibinfo {year} {2016})\BibitemShut {NoStop}%
\bibitem [{\citenamefont {Porter}\ \emph {et~al.}(2018)\citenamefont {Porter},
  \citenamefont {Rowell}, \citenamefont {J{\'{o}}hannesson},\ and\
  \citenamefont {Moskalenko}}]{Porter2018}%
  \BibitemOpen
  \bibfield  {author} {\bibinfo {author} {\bibfnamefont {T.}~\bibnamefont
  {Porter}}, \bibinfo {author} {\bibfnamefont {G.}~\bibnamefont {Rowell}},
  \bibinfo {author} {\bibfnamefont {G.}~\bibnamefont {J{\'{o}}hannesson}},\
  and\ \bibinfo {author} {\bibfnamefont {I.}~\bibnamefont {Moskalenko}},\
  }\bibfield  {journal} {\bibinfo  {journal} {Physical Review D}\ }\textbf
  {\bibinfo {volume} {98}},\ \href {https://doi.org/10.1103/physrevd.98.041302}
  {10.1103/physrevd.98.041302} (\bibinfo {year} {2018})\BibitemShut {NoStop}%
\bibitem [{\citenamefont {Griest}\ and\ \citenamefont
  {Kamionkowski}(1990)}]{Griest1990}%
  \BibitemOpen
  \bibfield  {author} {\bibinfo {author} {\bibfnamefont {K.}~\bibnamefont
  {Griest}}\ and\ \bibinfo {author} {\bibfnamefont {M.}~\bibnamefont
  {Kamionkowski}},\ }\href {https://doi.org/10.1103/physrevlett.64.615}
  {\bibfield  {journal} {\bibinfo  {journal} {Physical Review Letters}\
  }\textbf {\bibinfo {volume} {64}},\ \bibinfo {pages} {615} (\bibinfo {year}
  {1990})}\BibitemShut {NoStop}%
\bibitem [{\citenamefont {Abeysekara}\ \emph {et~al.}(2017)\citenamefont
  {Abeysekara}, \citenamefont {Albert}, \citenamefont {Alfaro}, \citenamefont
  {Alvarez}, \citenamefont {{\'{A}}lvarez},\ and\ \citenamefont
  {et~al.}}]{Abeysekara2017}%
  \BibitemOpen
  \bibfield  {author} {\bibinfo {author} {\bibfnamefont {A.~U.}\ \bibnamefont
  {Abeysekara}}, \bibinfo {author} {\bibfnamefont {A.}~\bibnamefont {Albert}},
  \bibinfo {author} {\bibfnamefont {R.}~\bibnamefont {Alfaro}}, \bibinfo
  {author} {\bibfnamefont {C.}~\bibnamefont {Alvarez}}, \bibinfo {author}
  {\bibfnamefont {J.~D.}\ \bibnamefont {{\'{A}}lvarez}},\ and\ \bibinfo
  {author} {\bibnamefont {et~al.}},\ }\href
  {https://doi.org/10.1126/science.aan4880} {\bibfield  {journal} {\bibinfo
  {journal} {Science}\ }\textbf {\bibinfo {volume} {358}},\ \bibinfo {pages}
  {911} (\bibinfo {year} {2017})}\BibitemShut {NoStop}%
\bibitem [{\citenamefont {J{\'{o}}hannesson}\ \emph {et~al.}(2019)\citenamefont
  {J{\'{o}}hannesson}, \citenamefont {Porter},\ and\ \citenamefont
  {Moskalenko}}]{Johannesson2019}%
  \BibitemOpen
  \bibfield  {author} {\bibinfo {author} {\bibfnamefont {G.}~\bibnamefont
  {J{\'{o}}hannesson}}, \bibinfo {author} {\bibfnamefont {T.~A.}\ \bibnamefont
  {Porter}},\ and\ \bibinfo {author} {\bibfnamefont {I.~V.}\ \bibnamefont
  {Moskalenko}},\ }\href {https://doi.org/10.3847/1538-4357/ab258e} {\bibfield
  {journal} {\bibinfo  {journal} {The Astrophysical Journal}\ }\textbf
  {\bibinfo {volume} {879}},\ \bibinfo {pages} {91} (\bibinfo {year}
  {2019})}\BibitemShut {NoStop}%
\bibitem [{\citenamefont {Albert}\ \emph {et~al.}(2021)\citenamefont {Albert},
  \citenamefont {Alfaro}, \citenamefont {Alvarez}, \citenamefont {Camacho},
  \citenamefont {Arteaga-Vel{\'{a}}zquez},\ and\ \citenamefont
  {et~al.}}]{Albert2021}%
  \BibitemOpen
  \bibfield  {author} {\bibinfo {author} {\bibfnamefont {A.}~\bibnamefont
  {Albert}}, \bibinfo {author} {\bibfnamefont {R.}~\bibnamefont {Alfaro}},
  \bibinfo {author} {\bibfnamefont {C.}~\bibnamefont {Alvarez}}, \bibinfo
  {author} {\bibfnamefont {J.~R.~A.}\ \bibnamefont {Camacho}}, \bibinfo
  {author} {\bibfnamefont {J.~C.}\ \bibnamefont {Arteaga-Vel{\'{a}}zquez}},\
  and\ \bibinfo {author} {\bibnamefont {et~al.}},\ }\href
  {https://doi.org/10.3847/2041-8213/abd77b} {\bibfield  {journal} {\bibinfo
  {journal} {The Astrophysical Journal}\ }\textbf {\bibinfo {volume} {907}},\
  \bibinfo {pages} {L30} (\bibinfo {year} {2021})}\BibitemShut {NoStop}%
\bibitem [{Note1()}]{Note1}%
  \BibitemOpen
  \bibinfo {note} {The only evident exception is the Cygnus Cocoon
  region}\BibitemShut {NoStop}%
\bibitem [{\citenamefont {Kulikov}\ and\ \citenamefont
  {Khristiansen}(1959)}]{Kulikov1959}%
  \BibitemOpen
  \bibfield  {author} {\bibinfo {author} {\bibfnamefont {G.}~\bibnamefont
  {Kulikov}}\ and\ \bibinfo {author} {\bibfnamefont {G.}~\bibnamefont
  {Khristiansen}},\ }\href@noop {} {\bibfield  {journal} {\bibinfo  {journal}
  {Sov. Physics JETP}\ }\textbf {\bibinfo {volume} {35}},\ \bibinfo {pages}
  {441} (\bibinfo {year} {1959})}\BibitemShut {NoStop}%
\bibitem [{\citenamefont {Fowler}\ \emph {et~al.}(2001)\citenamefont {Fowler},
  \citenamefont {Fortson}, \citenamefont {Jui}, \citenamefont {Kieda},
  \citenamefont {Ong}, \citenamefont {Pryke},\ and\ \citenamefont
  {Sommers}}]{Fowler2001}%
  \BibitemOpen
  \bibfield  {author} {\bibinfo {author} {\bibfnamefont {J.}~\bibnamefont
  {Fowler}}, \bibinfo {author} {\bibfnamefont {L.}~\bibnamefont {Fortson}},
  \bibinfo {author} {\bibfnamefont {C.}~\bibnamefont {Jui}}, \bibinfo {author}
  {\bibfnamefont {D.}~\bibnamefont {Kieda}}, \bibinfo {author} {\bibfnamefont
  {R.}~\bibnamefont {Ong}}, \bibinfo {author} {\bibfnamefont {C.}~\bibnamefont
  {Pryke}},\ and\ \bibinfo {author} {\bibfnamefont {P.}~\bibnamefont
  {Sommers}},\ }\href {https://doi.org/10.1016/s0927-6505(00)00139-0}
  {\bibfield  {journal} {\bibinfo  {journal} {Astroparticle Physics}\ }\textbf
  {\bibinfo {volume} {15}},\ \bibinfo {pages} {49} (\bibinfo {year}
  {2001})}\BibitemShut {NoStop}%
\bibitem [{\citenamefont {Aglietta}\ \emph {et~al.}(2004)\citenamefont
  {Aglietta}, \citenamefont {Alessandro}, \citenamefont {Antonioli},
  \citenamefont {Arneodo}, \citenamefont {Bergamasco},\ and\ \citenamefont
  {et~al.}}]{Aglietta2004}%
  \BibitemOpen
  \bibfield  {author} {\bibinfo {author} {\bibfnamefont {M.}~\bibnamefont
  {Aglietta}}, \bibinfo {author} {\bibfnamefont {B.}~\bibnamefont
  {Alessandro}}, \bibinfo {author} {\bibfnamefont {P.}~\bibnamefont
  {Antonioli}}, \bibinfo {author} {\bibfnamefont {F.}~\bibnamefont {Arneodo}},
  \bibinfo {author} {\bibfnamefont {L.}~\bibnamefont {Bergamasco}},\ and\
  \bibinfo {author} {\bibnamefont {et~al.}},\ }\href
  {https://doi.org/10.1016/j.astropartphys.2004.04.005} {\bibfield  {journal}
  {\bibinfo  {journal} {Astroparticle Physics}\ }\textbf {\bibinfo {volume}
  {21}},\ \bibinfo {pages} {583} (\bibinfo {year} {2004})}\BibitemShut
  {NoStop}%
\bibitem [{\citenamefont {Antoni}\ \emph {et~al.}(2005)\citenamefont {Antoni},
  \citenamefont {Apel}, \citenamefont {Badea}, \citenamefont {Bekk},
  \citenamefont {Bercuci},\ and\ \citenamefont {et~al.}}]{Antoni2005}%
  \BibitemOpen
  \bibfield  {author} {\bibinfo {author} {\bibfnamefont {T.}~\bibnamefont
  {Antoni}}, \bibinfo {author} {\bibfnamefont {W.}~\bibnamefont {Apel}},
  \bibinfo {author} {\bibfnamefont {A.}~\bibnamefont {Badea}}, \bibinfo
  {author} {\bibfnamefont {K.}~\bibnamefont {Bekk}}, \bibinfo {author}
  {\bibfnamefont {A.}~\bibnamefont {Bercuci}},\ and\ \bibinfo {author}
  {\bibnamefont {et~al.}},\ }\href
  {https://doi.org/10.1016/j.astropartphys.2005.04.001} {\bibfield  {journal}
  {\bibinfo  {journal} {Astroparticle Physics}\ }\textbf {\bibinfo {volume}
  {24}},\ \bibinfo {pages} {1} (\bibinfo {year} {2005})}\BibitemShut {NoStop}%
\bibitem [{\citenamefont {Aartsen}\ \emph {et~al.}(2013)\citenamefont
  {Aartsen}, \citenamefont {Abbasi}, \citenamefont {Abdou}, \citenamefont
  {Ackermann}, \citenamefont {Adams},\ and\ \citenamefont
  {et~al.}}]{Aartsen2013}%
  \BibitemOpen
  \bibfield  {author} {\bibinfo {author} {\bibfnamefont {M.~G.}\ \bibnamefont
  {Aartsen}}, \bibinfo {author} {\bibfnamefont {R.}~\bibnamefont {Abbasi}},
  \bibinfo {author} {\bibfnamefont {Y.}~\bibnamefont {Abdou}}, \bibinfo
  {author} {\bibfnamefont {M.}~\bibnamefont {Ackermann}}, \bibinfo {author}
  {\bibfnamefont {J.}~\bibnamefont {Adams}},\ and\ \bibinfo {author}
  {\bibnamefont {et~al.}},\ }\bibfield  {journal} {\bibinfo  {journal}
  {Physical Review D}\ }\textbf {\bibinfo {volume} {88}},\ \href
  {https://doi.org/10.1103/physrevd.88.042004} {10.1103/physrevd.88.042004}
  (\bibinfo {year} {2013})\BibitemShut {NoStop}%
\bibitem [{\citenamefont {Montmerle}(1979)}]{Montmerle1979}%
  \BibitemOpen
  \bibfield  {author} {\bibinfo {author} {\bibfnamefont {T.}~\bibnamefont
  {Montmerle}},\ }\href {https://doi.org/10.1086/157166} {\bibfield  {journal}
  {\bibinfo  {journal} {The Astrophysical Journal}\ }\textbf {\bibinfo {volume}
  {231}},\ \bibinfo {pages} {95} (\bibinfo {year} {1979})}\BibitemShut
  {NoStop}%
\bibitem [{\citenamefont {Casse}\ and\ \citenamefont {Paul}(1980)}]{Casse1980}%
  \BibitemOpen
  \bibfield  {author} {\bibinfo {author} {\bibfnamefont {M.}~\bibnamefont
  {Casse}}\ and\ \bibinfo {author} {\bibfnamefont {J.~A.}\ \bibnamefont
  {Paul}},\ }\href {https://doi.org/10.1086/157863} {\bibfield  {journal}
  {\bibinfo  {journal} {The Astrophysical Journal}\ }\textbf {\bibinfo {volume}
  {237}},\ \bibinfo {pages} {236} (\bibinfo {year} {1980})}\BibitemShut
  {NoStop}%
\bibitem [{\citenamefont {Cesarsky}\ and\ \citenamefont
  {Montmerle}(1983)}]{Cesarsky1983}%
  \BibitemOpen
  \bibfield  {author} {\bibinfo {author} {\bibfnamefont {C.~J.}\ \bibnamefont
  {Cesarsky}}\ and\ \bibinfo {author} {\bibfnamefont {T.}~\bibnamefont
  {Montmerle}},\ }\href {https://doi.org/10.1007/bf00167503} {\bibfield
  {journal} {\bibinfo  {journal} {Space Science Reviews}\ }\textbf {\bibinfo
  {volume} {36}},\ \bibinfo {pages} {173} (\bibinfo {year} {1983})}\BibitemShut
  {NoStop}%
\bibitem [{\citenamefont {Bykov}(2014)}]{Bykov2014}%
  \BibitemOpen
  \bibfield  {author} {\bibinfo {author} {\bibfnamefont {A.~M.}\ \bibnamefont
  {Bykov}},\ }\bibfield  {journal} {\bibinfo  {journal} {The Astronomy and
  Astrophysics Review}\ }\textbf {\bibinfo {volume} {22}},\ \href
  {https://doi.org/10.1007/s00159-014-0077-8} {10.1007/s00159-014-0077-8}
  (\bibinfo {year} {2014})\BibitemShut {NoStop}%
\bibitem [{\citenamefont {Aharonian}\ \emph {et~al.}(2019)\citenamefont
  {Aharonian}, \citenamefont {Yang},\ and\ \citenamefont
  {de~O{\~{n}}a~Wilhelmi}}]{Aharonian2019}%
  \BibitemOpen
  \bibfield  {author} {\bibinfo {author} {\bibfnamefont {F.}~\bibnamefont
  {Aharonian}}, \bibinfo {author} {\bibfnamefont {R.}~\bibnamefont {Yang}},\
  and\ \bibinfo {author} {\bibfnamefont {E.}~\bibnamefont
  {de~O{\~{n}}a~Wilhelmi}},\ }\href {https://doi.org/10.1038/s41550-019-0724-0}
  {\bibfield  {journal} {\bibinfo  {journal} {Nature Astronomy}\ }\textbf
  {\bibinfo {volume} {3}},\ \bibinfo {pages} {561} (\bibinfo {year}
  {2019})}\BibitemShut {NoStop}%
\bibitem [{\citenamefont {Lipari}(2019)}]{Lipari2019}%
  \BibitemOpen
  \bibfield  {author} {\bibinfo {author} {\bibfnamefont {P.}~\bibnamefont
  {Lipari}},\ }\bibfield  {journal} {\bibinfo  {journal} {Physical Review D}\
  }\textbf {\bibinfo {volume} {99}},\ \href
  {https://doi.org/10.1103/physrevd.99.043005} {10.1103/physrevd.99.043005}
  (\bibinfo {year} {2019})\BibitemShut {NoStop}%
\bibitem [{\citenamefont {H\"{o}randel}(2003)}]{Hoerandel2003}%
  \BibitemOpen
  \bibfield  {author} {\bibinfo {author} {\bibfnamefont {J.~R.}\ \bibnamefont
  {H\"{o}randel}},\ }\href {https://doi.org/10.1016/s0927-6505(02)00198-6}
  {\bibfield  {journal} {\bibinfo  {journal} {Astroparticle Physics}\ }\textbf
  {\bibinfo {volume} {19}},\ \bibinfo {pages} {193} (\bibinfo {year}
  {2003})}\BibitemShut {NoStop}%
\bibitem [{\citenamefont {Guo}\ \emph {et~al.}(2014)\citenamefont {Guo},
  \citenamefont {Hu}, \citenamefont {Yuan}, \citenamefont {Tian},\ and\
  \citenamefont {Gao}}]{Guo2014}%
  \BibitemOpen
  \bibfield  {author} {\bibinfo {author} {\bibfnamefont {Y.~Q.}\ \bibnamefont
  {Guo}}, \bibinfo {author} {\bibfnamefont {H.~B.}\ \bibnamefont {Hu}},
  \bibinfo {author} {\bibfnamefont {Q.}~\bibnamefont {Yuan}}, \bibinfo {author}
  {\bibfnamefont {Z.}~\bibnamefont {Tian}},\ and\ \bibinfo {author}
  {\bibfnamefont {X.~J.}\ \bibnamefont {Gao}},\ }\href
  {https://doi.org/10.1088/0004-637x/795/1/100} {\bibfield  {journal} {\bibinfo
   {journal} {The Astrophysical Journal}\ }\textbf {\bibinfo {volume} {795}},\
  \bibinfo {pages} {100} (\bibinfo {year} {2014})}\BibitemShut {NoStop}%
\bibitem [{\citenamefont {Bai}\ \emph {et~al.}(2019)\citenamefont {Bai},
  \citenamefont {Bi}, \citenamefont {Bi}, \citenamefont {Cao}, \citenamefont
  {Chen},\ and\ \citenamefont {et~al.}}]{LHAASO-arxiv-2019}%
  \BibitemOpen
  \bibfield  {author} {\bibinfo {author} {\bibfnamefont {X.}~\bibnamefont
  {Bai}}, \bibinfo {author} {\bibfnamefont {B.~Y.}\ \bibnamefont {Bi}},
  \bibinfo {author} {\bibfnamefont {X.~J.}\ \bibnamefont {Bi}}, \bibinfo
  {author} {\bibfnamefont {Z.}~\bibnamefont {Cao}}, \bibinfo {author}
  {\bibfnamefont {S.~Z.}\ \bibnamefont {Chen}},\ and\ \bibinfo {author}
  {\bibnamefont {et~al.}},\ }\href@noop {} {\bibinfo {title} {The large high
  altitude air shower observatory (lhaaso) science white paper}} (\bibinfo
  {year} {2019}),\ \Eprint {https://arxiv.org/abs/1905.02773} {arXiv:1905.02773
  [astro-ph.HE]} \BibitemShut {NoStop}%
\bibitem [{\citenamefont {Actis}\ \emph {et~al.}(2011)\citenamefont {Actis},
  \citenamefont {Agnetta}, \citenamefont {Aharonian}, \citenamefont {A},
  \citenamefont {Aleksic},\ and\ \citenamefont {et~al.}}]{Actis2011}%
  \BibitemOpen
  \bibfield  {author} {\bibinfo {author} {\bibfnamefont {M.}~\bibnamefont
  {Actis}}, \bibinfo {author} {\bibfnamefont {G.}~\bibnamefont {Agnetta}},
  \bibinfo {author} {\bibfnamefont {F.}~\bibnamefont {Aharonian}}, \bibinfo
  {author} {\bibfnamefont {A.}~\bibnamefont {A}}, \bibinfo {author}
  {\bibfnamefont {J.}~\bibnamefont {Aleksic}},\ and\ \bibinfo {author}
  {\bibnamefont {et~al.}},\ }\href {https://doi.org/10.1007/s10686-011-9247-0}
  {\bibfield  {journal} {\bibinfo  {journal} {Experimental Astronomy}\ }\textbf
  {\bibinfo {volume} {32}},\ \bibinfo {pages} {193} (\bibinfo {year}
  {2011})}\BibitemShut {NoStop}%
\bibitem [{\citenamefont {Brun}\ and\ \citenamefont
  {Rademakers}(1997)}]{Brun1997}%
  \BibitemOpen
  \bibfield  {author} {\bibinfo {author} {\bibfnamefont {R.}~\bibnamefont
  {Brun}}\ and\ \bibinfo {author} {\bibfnamefont {F.}~\bibnamefont
  {Rademakers}},\ }\href {https://doi.org/10.1016/s0168-9002(97)00048-x}
  {\bibfield  {journal} {\bibinfo  {journal} {NIM A}\ }\textbf {\bibinfo
  {volume} {389}},\ \bibinfo {pages} {81} (\bibinfo {year} {1997})}\BibitemShut
  {NoStop}%
\end{thebibliography}%
\end{document}